\documentclass[%
reprint,
superscriptaddress,
amsmath,amssymb,
aps,98
]{revtex4-2}

\usepackage [pdftex] {graphicx}
\usepackage {epstopdf}

\usepackage{dcolumn}
\usepackage{bm}
\usepackage[utf8]{inputenc}
\usepackage{color,xcolor}

\usepackage{subfigure}
\usepackage{caption}
\usepackage{url}
\usepackage{hyperref}
\usepackage{booktabs}

\usepackage{amsmath, graphicx,algorithm,algorithmic,amsfonts,float,multirow}

\begin{document}
	
\preprint{APS/123-QED}

\title{Identification of Stochastic Gravitational Wave Backgrounds from Cosmic String Using Machine Learning }

\author{Xianghe Ma}
\affiliation{%
    Department of Physics,Chongqing University, 
Chongqing 401331, P.R. China\\
}%
 \affiliation{%
    Chongqing Key Laboratory for Strongly Coupled Physics, Chongqing University, Chongqing 401331, P.R. China\\
}%

\author{Borui Wang}%
\affiliation{%
    Department of Earth and Sciences, Southern University of Science and Technology, 
Shenzhen 518055, P.R. China\\
}%

\author{Nan Yang}%
\affiliation{%
    Department of Electronical Information Science and Technology, Xingtai University, Xingtai 054001, China\\
}%

\author{Jin Li}%
\email{cqujinli1983@cqu.edu.cn}
\affiliation{%
    Department of Physics,Chongqing University, 
Chongqing 401331, P.R. China\\
}%
\affiliation{%
Chongqing Key Laboratory for Strongly Coupled Physics, Chongqing University, Chongqing 401331, P.R. China\\
}%

\author{Brendan McCane}%
\affiliation{%
    School of Computing, University of Otago, Otago 9016, New Zealand\\
}%

\author{Mengfei Sun}%
\affiliation{%
    Department of Physics,Chongqing University, 
Chongqing 401331, P.R. China\\
}%
\affiliation{%
Chongqing Key Laboratory for Strongly Coupled Physics, Chongqing University, Chongqing 401331, P.R. China\\
}%

\author{Jie Wu}%
\affiliation{%
    Department of Physics,Chongqing University, 
Chongqing 401331, P.R. China\\
}%
\affiliation{%
Chongqing Key Laboratory for Strongly Coupled Physics, Chongqing University, Chongqing 401331, P.R. China\\
}%

\author{Minghui Zhang}%
\affiliation{%
    Department of Physics, Southern University of Science and Technology, 
Shenzhen 518055, P.R. China\\
}%

\author{Yan Meng}%
\email{200221007@xttc.edu.cn}
\affiliation{%
    Department of Physics,Chongqing University, 
Chongqing 401331, P.R. China\\
}%
\affiliation{%
Department of Physics, Xingtai University, Xingtai 054001, China\\
}%

\date{\today}

\begin{abstract}  
Cosmic strings play a crucial role in enhancing our understanding of the fundamental structure and evolution of the universe, unifying our knowledge of cosmology, and potentially unveiling new physical laws and phenomena. The advent and operation of space-based detectors provide an important opportunity for detecting stochastic gravitational wave backgrounds (SGWB) generated by cosmic strings. However, the intricate nature of SGWB poses a formidable challenge in distinguishing its signal from the complex noise by some traditional methods. Therefore, we attempt to identify SGWB based on machine learning. Our findings show that the joint detection of LISA and Taiji significantly outperforms individual detectors, and even in the presence of numerous low signal-to-noise ratio(SNR) signals, the identification accuracy remains exceptionally high with 95\%. Although our discussion is based solely on simulated data, the relevant methods can provide data-driven analytical capabilities for future observations of SGWB.

\end{abstract}

\maketitle

\section{Introduction}
The LIGO/Virgo collaboration made a groundbreaking announcement with the first direct detection of gravitational waves, produced by the merger of binary black holes (BBHs) \cite{abbott2016observation}. This discovery not only confirmed Einstein's general theory of relativity but also marked the dawn of gravitational wave astronomy. With continuous advancements in detector sensitivity, the number of observed gravitational wave events has steadily increased. Since the first detection in 2015 (GW150914), ground-based detectors have successfully recorded over 100 events, primarily in the frequency range of tens to hundreds of hertz. However, a significant gap still exists in the detection of low-frequency gravitational waves. To fill this gap, space-based gravitational wave detectors like LISA \cite{amaro2017laser}, the Taiji program \cite{luo2020brief}, and the TianQin project \cite{luo2016tianqin} are being developed, offering new opportunities for low-frequency gravitational wave detection.

Space-based detectors are not only capable of detecting a wide range of compact binary coalescence (CBC) events \cite{pitkin2011gravitational}, but they also serve as powerful tools for investigating stochastic gravitational wave backgrounds (SGWB) in the millihertz frequency band. The SGWB consists of a superposition of numerous independent gravitational wave signals from various directions in the sky, originating from a variety of sources across the universe. These sources are classified into cosmological and astrophysical categories, spanning different epochs in the history of the universe.

The cosmological sources of the SGWB include primordial processes in the early universe, such as inflation \cite{turner1997detectability} \cite{guzzetti2016gravitational}, cosmic strings \cite{sarangi2002cosmic,damour2000gravitational,stott2017gravitational,siemens2007gravitational}, and phase transitions \cite{witten1984cosmic,kosowsky1992gravitational,dev2016probing,von2020peccei}. This paper primarily focuses on the SGWB generated by cosmic strings. The study of SGWB is crucial as it provides unique insights into the early universe, fundamental physics, and astrophysical populations and processes. Detecting and characterizing the SGWB can help validate inflation models, explore the physics of the early universe, and illuminate the population and evolution of compact objects over cosmic time \cite{caprini2018cosmological,kuroyanagi2018probing,christensen2018stochastic}.

Due to the inherent noise characteristics of gravitational wave detectors \cite{phinney2002lisa} \cite{liu2023confusion}, particularly the prevalent foreground noise in space-based detectors, identifying gravitational wave signals from the data is a significant challenge. Traditional signal identification methods, such as Bayesian inference \cite{box2011bayesian} \cite{von2011bayesian} and matched filtering \cite{turin1960introduction} \cite{stankovic2023convolutional}, have limitations when applied to the detection of SGWB. For example, Bayesian inference requires sufficient prior information, which is difficult to obtain for SGWB, and it involves considerable offline processing time. Matched filtering relies on correlating the received signal with known target templates; however, the stochastic gravitational wave background lacks defined waveform templates. As an alternative, we employ machine learning techniques to identify SGWB, offering a promising approach that complements traditional methods.

Machine learning has made significant progress in various gravitational wave (GW) data analysis tasks, including GW detection \cite{george2018deep, gabbard2018matching, zhang2022detecting, lopez2021deep}, parameter estimation \cite{gabbard2022bayesian} \cite{dax2021real}, glitch classification \cite{colgan2020efficient, cavaglia2018finding, razzano2018image}, noise reduction \cite{ormiston2020noise} \cite{mogushi2021reduction}, and signal extraction \cite{torres2016denoising, wei2020gravitational, shen2017denoising, chatterjee2021extraction}. In Ref. \cite{zhao2023space}, Zhoujian Cao et al. applied machine learning techniques to identify SGWB within a general model and achieved promising results. In contrast, this study focuses on the SGWB generated by cosmic strings, considering a broader signal-to-noise ratio(SNR) range than the work in \cite{zhao2023space}, and utilizes multi-detector joint observations with foreground noise.

This paper is organized as follows: Section 2 describes the data simulation process, including the simulation of SGWB signals, the sensitivity curves for LISA and Taiji, and the foreground noise. Section 3 provides a brief introduction to the machine learning network used in this study and discusses some key hyperparameters. Section 4 presents the results and analysis, including outcomes with detector noise and foreground noise. Finally, Section 5 concludes the paper and offers some remarks.

\section{Data simulation}
Our simulated data are time-domain sequences consisting of six groups. The six groups refer to different kinds of noise (${n_i}(t)$, i=1,2,3,4,5,6). In each group, the samples can be divided into two categories: one containing the SGWB from cosmic strings with noise, i.e., ${h_{SGWB}} + {n_i}(t)$; and the other only containing noise i.e., ${n_i}(t)$. ${n_{i = 1,2,3}}(t)$ stands for noise, respectively:
\begin{itemize}
    \item ${n_1}(t)$ is the instrumental noise from LISA detector,
    \item ${n_2}(t)$ is the instrumental noise from Taiji detector, 
    \item ${n_3}(t)$ is the instrumental noise from LISA and Taiji detectors. 
\end{itemize}
The third group ${h_{SGWB}} + {n_3}(t)$ combines the data from the first two groups to simulate the effect of joint observation by LISA and Taiji. The remaining three groups are based on the joint observation instrumental noise with the additional foreground noises:
\begin{itemize}
    \item ${n_4}(t)$ is ${n_3}(t)$ plus the foreground noise from double white dwarfs,
    \item ${n_5}(t)$ is ${n_3}(t)$ plus gravitational wave background noise generated by binary black holes (BBH) and binary neutron stars (BNS) observed by LIGO and Virgo, 
    \item ${n_6}(t)$ is cosist of ${n_4}(t) + {n_5}(t)$.  
\end{itemize}

\subsection{The stochastic gravitational wave background signals generated by cosmic strings}
We adopt the model proposed in~\cite{wang2024ability,wang2023probing,sousa2020full} to describe the gravitational waves generated by cosmic strings, employing highly accurate analytical approximation formulas for their representation~\cite{blanco2014number}\cite{ringeval2007cosmological}. The gravitational wave signals are influenced significantly by the cosmic string tension $Gu$, which characterizes the size of the loops. The value of $Gu$ ranges from \( 10^{-9} \) to \( 10^{-7} \), and we adopt a uniform distribution for sampling within each order of magnitude. We introduce a free parameter \( \alpha \) to denote the string loop size and define the total power emitted by the cosmic strings as \( \Gamma = 50 \). The value of \( \alpha \) ranges from \( 10^{-3} \) to \( 10^{0} \), which are also sampled with a uniform distribution.

For the SGWB from cosmic strings, the contributions from cosmic string loops can be categorized into three periods: 
\begin{itemize}
    \item loops formed and decayed during the radiation era,
    \item loops formed during the radiation era and decayed during the matter era,  
    \item loops formed during the matter era.
\end{itemize}
For loops formed and decayed in the radiation region, the form of stochastic gravitational wave background is given by 
\begin{equation}
\Omega_{GW}^{r}\left( f \right) = \frac{128}{9}\pi A_{r}\Omega_{r}\frac{Gu}{\varepsilon_{r}} 
\left[ \left( \frac{f\left( 1+\varepsilon_{r} \right)}{\frac{B_{r}\Omega_{m}}{\Omega_{r}}+f} \right)^{\frac{3}{2}} - 1 \right],
\label{1}
\end{equation}
here
\begin{align}
\varepsilon_{\text{r}} = \frac{\alpha }{\Gamma G\mu}, \quad
A_{r} = \frac{\tilde{c}}{\sqrt{2}}F\frac{v_{r}}{\xi_{r}^{3}}, \quad
B_{r} = \frac{2H_{0}\Omega_{r}^{\frac{1}{2}}}{\nu_{r}\Gamma G\mu},
\end{align}
The label $r$ indicates the radiation era and in these equations ${{v}_{r}}=0.662$, ${{\xi }_{r}}=0.271$, ${{\nu }_{r}}=1/2$, $F=0.1$ and $\tilde{c}$ is a phenomenological parameter which can be set as $\tilde{c}=0.23\pm 0.04$~\cite{martins2002extending}. In our work, the evolution of the universe is assumed to follow a standard $\Lambda$CDM model, with its underlying parameters are ${{H}_{0}}=100h$ $km/\left( s\cdot Mpc \right)$, $h=0.678$, ${{\Omega }_{r}}=8.397\times {{10}^{-5}}$, ${{\Omega }_{m}}=0.308$.
For loops formed in the radiation region and decayed in the matter region, their contribution to SGWB has the following form
\begin{align}
\Omega_{GW}^{r\text{m}}(f) &= 32\sqrt{3}\pi \left({\Omega_{m}\Omega_{r}}\right)^{\frac{3}{4}} H_{0} \frac{A_{r}}{\Gamma} 
\frac{\left(1+\varepsilon_{r}\right)^{\frac{3}{2}}}{f^{\frac{1}{2}}\varepsilon_{r}} \notag \\
&\quad \times \left\{ 
\frac{\left(\frac{\Omega_{m}}{\Omega_{r}}\right)^{\frac{1}{4}}}
{\left(B_{m}\left(\frac{\Omega_{m}}{\Omega_{r}}\right)^{\frac{1}{2}} + f\right)^{\frac{1}{2}}}
\left[ 2 + \frac{f}{B_{m}\left(\frac{\Omega_{m}}{\Omega_{r}}\right)^{\frac{1}{2}} + f} \right] \right. \notag \\
&\quad \left. - \frac{1}{B_{m} + f} 
\left[ 2 + \frac{f}{B_{m} + f} \right] \right\},
\end{align}
here
\begin{align}
{{B}_{r}}=\frac{2{{H}_{0}}\Omega _{m}^{\frac{1}{2}}}{{{\nu }_{m}}\Gamma Gu},
\end{align}
where ${{\nu }_{m}}=2/3$. The contribution of loops generated in the matter period to the SGWB generation by cosmic strings is given by
\begin{align}
\Omega_{GW}^{m}(f) &= 54\pi H_{0} \Omega_{m}^{\frac{3}{2}} \frac{A_{m}}{\Gamma} \frac{\varepsilon_{m} + 1}{\varepsilon_{m}} \frac{B_{m}}{f} \notag \\
&\times \left\{ \frac{2B_{m} + f}{B_{m}(B_{m} + f)} - \frac{1}{f} \frac{2\varepsilon_{m} + 1}{\varepsilon_{m}(\varepsilon_{m} + 1)} \right. \notag \\
&\quad \left. + \frac{2}{f} \log \left( \frac{\varepsilon_{m} + 1}{\varepsilon_{m}} \frac{B_{m}}{B_{m} + f} \right) \right\},
\end{align}
here
\begin{align}
{{\varepsilon }_{m}}=\frac{\alpha }{\Gamma Gu},{{A}_{m}}=\frac{{\tilde{c}}}{\sqrt{2}}F\frac{{{v}_{m}}}{{{\xi }_{m}}^{3}}.
\end{align}
The label $m$ indicates the matter era and in these equations ${{v}_{m}}=0.583$, ${{\xi }_{m}}=0.625$. Therefore, the SGWB generated by cosmic strings can be well approximated as
\begin{align}
{{\Omega }_{GW}}\left( f \right)=\Omega _{GW}^{r}\left( f \right)+\Omega _{GW}^{rm}\left( f \right)+\Omega _{GW}^{m}\left( f \right).
\end{align}

For the dimensionless energy density in GWs ${{\Omega }_{GW}}$, we need to convert to the frequency domain strain $h\left( f \right)$~\cite{moore2014gravitational}, given by
\begin{align}
H_0^2{\Omega _{GW}}\left( f \right) = \frac{{8{\pi ^2}}}{3}{f^4}{\left| {h\left( f \right)} \right|^2}.
\end{align}

We perform a Fourier transform and add random phases to convert the frequency domain signal to a time domain signal.

\subsection{Simulation of instrument noise}
Leveraging the triangular geometry of LISA-like detectors, time-delay interferometry (TDI) techniques can be employed to combine phase differences with varying time delays, effectively canceling laser frequency noise~\cite{boileau2023prospects, wang2021alternative}. For simplicity, we assume that the instrument noise consists of two primary components: test mass acceleration noise and optical path length fluctuations. These noise sources are considered identical for each spacecraft. Given that the arm lengths are equal, the LISA instrument effectively forms an equilateral triangle~\cite{vallisneri2012non}.

For computational convenience, we adopt approximations of the gravitational wave (GW) response in the A, E, and T channels, as provided in~\cite{smith2019lisa}:
\begin{align}
R_A^i\left( f \right) = R_E^i\left( f \right) = \frac{9}{{20}}{\left| {{W^i}\left( f \right)} \right|^2}{\left[ {1 + {{\left( {\frac{f}{{4{f_i}/3}}} \right)}^2}} \right]^{ - 1}},
\end{align}
and
\begin{align}
R_T^i\left( f \right) \simeq \frac{1}{{4032}}{\left( {\frac{f}{{{f_i}}}} \right)^6}{\left| {{W^i}\left( f \right)} \right|^2}{\left[ {1 + \frac{5}{{16128}}{{\left( {\frac{f}{{{f_i}}}} \right)}^8}} \right]^{ - 1}},
\end{align}
where $i = LISA,Taiji$, ${W^i}\left( f \right) = 1 - {e^{ - 2if/{f_i}}}$, and for the LISA-like detector, ${f_i} = c/\left( {2\pi {L_i}} \right)$, with ${L_{LISA}} = 2.5 \times {10^6}$ $km$, and ${L_{Taiji}} = 3 \times {10^6}$  $km$.

We adopt the noise model outlined in the LISA Science Requirements Document~\cite{phinney2002lisa, baker2019laser} and assume that the same model applies to the LISA-like detector, Taiji. This model presumes that the noise in all channels remains constant and identical~\cite{wang2021alternative}, specifically accounting for two primary noise sources: acceleration noise and optical path disturbance noise, which are explicitly defined as follows~\cite{amaro2017laser, hu2017taiji}:
\begin{align}
N_{acc}^i\left( f \right) &= \frac{{N_a^i}}{{{{\left( {2\pi f} \right)}^4}}}\left( {1 + {{\left( {\frac{{f_1}}{f}} \right)}^2}} \right) \notag \\
&= \frac{{\left( {\sqrt {{{\left( {\delta {a_i}} \right)}^2}} /{L_i}} \right)}}{{{{\left( {2\pi f} \right)}^4}}}\left( {1 + {{\left( {\frac{{{f_1}}}{f}} \right)}^2}} \right),
\end{align}
\begin{align}
N_{op}^i\left( f \right) = N_o^i = {\left( {\sqrt {{{\left( {\delta {x_i}} \right)}^2}} /{L_i}} \right)^2},
\end{align}
where
\begin{align}
\sqrt {{{\left( {\delta {a_{LISA}}} \right)}^2}}  = 3 \times {10^{ - 15}}  m/{s^2}, \\
\sqrt {{{\left( {\delta {x_{LISA}}} \right)}^2}}  = 1.5 \times {10^{ - 11}}  m,
\end{align}
and~\cite{ruan2020lisa}
\begin{align}
\sqrt {{{\left( {\delta {a_{Taiji}}} \right)}^2}}  = 3 \times {10^{ - 15}}  m/{s^2}, \\
\sqrt {{{\left( {\delta {x_{Taiji}}} \right)}^2}}  = 8 \times {10^{ - 12}}  m,
\end{align}
where $i = LISA,Taiji$, ${f_1} = 0.4$ $mHz$. These noise models can be transformed into interferometer noise through
\begin{align}
N_X^i\left( f \right) = \left[ {4N_{op}^i\left( f \right) + 8\left[ {1 + {{\cos }^2}\left( {f/{f_i}} \right)} \right]N_{acc}^i\left( f \right)} \right]{\left| {{W^i}\left( f \right)} \right|^2},
\end{align}
\begin{align}
N_{XY}^i\left( f \right) =  - \left[ {2N_{op}^i\left( f \right) + 8N_{acc}^i\left( f \right)} \right]\cos \left( {f/{f_i}} \right){\left| {{W^i}\left( f \right)} \right|^2}.
\end{align}
The noise models for the power spectral densities in the AET channels are obtained by diagonalizing the covariance matrix of the XYZ channels. The resulting diagonal entries are given by:
\begin{align}
N_A^i\left( f \right) = N_E^i\left( f \right) = N_X^i\left( f \right) - N_{XY}^i\left( f \right),
\end{align}
\begin{align}
N_T^i\left( f \right) = N_X^i\left( f \right) + 2N_{XY}^i\left( f \right).
\end{align}

The noise spectral density for different channels can be derived from the noise power spectral density and the corresponding response function, expressed as:
\begin{align}
S_j^i\left( f \right) = \frac{{N_j^i\left( f \right)}}{{R_j^i\left( f \right)}},
\end{align}
where $j = A,E,T$. Based on the noise spectral density, the total equivalent energy density for a single LISA-like detector can be formulated as follows~\cite{smith2019lisa}:
\begin{align}
{\Omega ^i} = \frac{{4{\pi ^2}{f^3}}}{{3H_0^2}}{\left( {\sum\limits_{j = A,E,T} {\frac{{R_j^i\left( f \right)}}{{N_j^i\left( f \right)}}} } \right)^{ - 1}}.
\end{align}

\subsection{Simulation of foreground noise}
In actual observations, the data consists not only of gravitational wave signals generated by cosmic strings but also of astrophysical foreground noise. This foreground noise primarily includes two components: double white dwarf (DWD) and inspiral binary black holes (BBH)/binary neutron stars (BNS) based on the observations of LIGO and Virgo~\cite{chen2019stochastic,boileau2022ability,wang2021algorithm}.

The gravitational wave model for double white dwarfs is a modulated signal based on the LISA orbital motion. Its energy spectral density can be approximated using the broken power-law model proposed in Refs.~\cite{boileau2023prospects, boileau2022ability}, which is given by:
\begin{align}
{\Omega _{DWD}}\left( f \right) = \frac{{{A_1}{{\left( {\frac{f}{{{f_i}}}} \right)}^{{\alpha _1}}}}}{{1 + {A_2}{{\left( {\frac{f}{{{f_i}}}} \right)}^{{\alpha _2}}}}},
\end{align}
where ${A_1} = 7.44 \times {10^{ - 14}}$, ${A_2} = 2.96 \times {10^{ - 7}}$, ${\alpha _1} =  - 1.98$, ${\alpha _2} =  - 2.6$. 

For the superposition of gravitational wave background produced by inspiralling BBHs/BNS observed by LIGO and Virgo, it is given by
\begin{align}
{\Omega _{astro}}\left( f \right) = {\Omega _{astro}}{\left( {\frac{f}{{{f_ * }}}} \right)^{{\alpha _{astro}}}},
\end{align}
where ${f_ * } = 3mHz$, ${\Omega _{astro}} = 4.44 \times {10^{ - 12}}$ and ${\alpha _{astro}} = 2/3$.
\begin{figure}[htbp]
	\centering
	\includegraphics[width=0.5\textwidth]{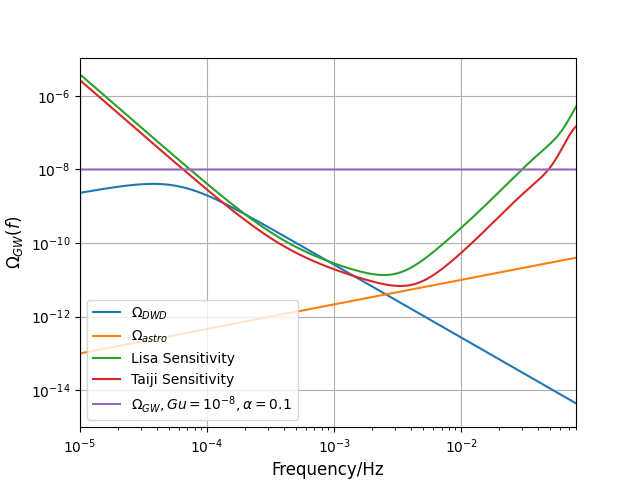}    
	\caption{\footnotesize
 The equivalent energy density of LISA, Taiji, foreground noise, SGWB. 
    }
\end{figure}

In Figure 1, we present the sensitivity curves of the detectors, the foreground noise, and the relationship between the equivalent energy density of the SGWB. The equivalent energy density of the noise is converted into power spectral density (PSD), and Gaussian noise in the time domain is generated to simulate both instrumental and foreground noise. The equivalent energy density of the DWD signal is determined based on the arm length of the LISA detector.

The sensitivity curves of LISA and Taiji are quite similar; however, Taiji's longer arm length results in lower instrumental noise. In the joint observation of LISA and Taiji, SGWB signals produced by cosmic strings are assumed to be long-lasting. Therefore, it is unlikely that one detector would observe the signal while the other does not.
\begin{figure}[h!]
    \centering
    \includegraphics[scale=0.2]{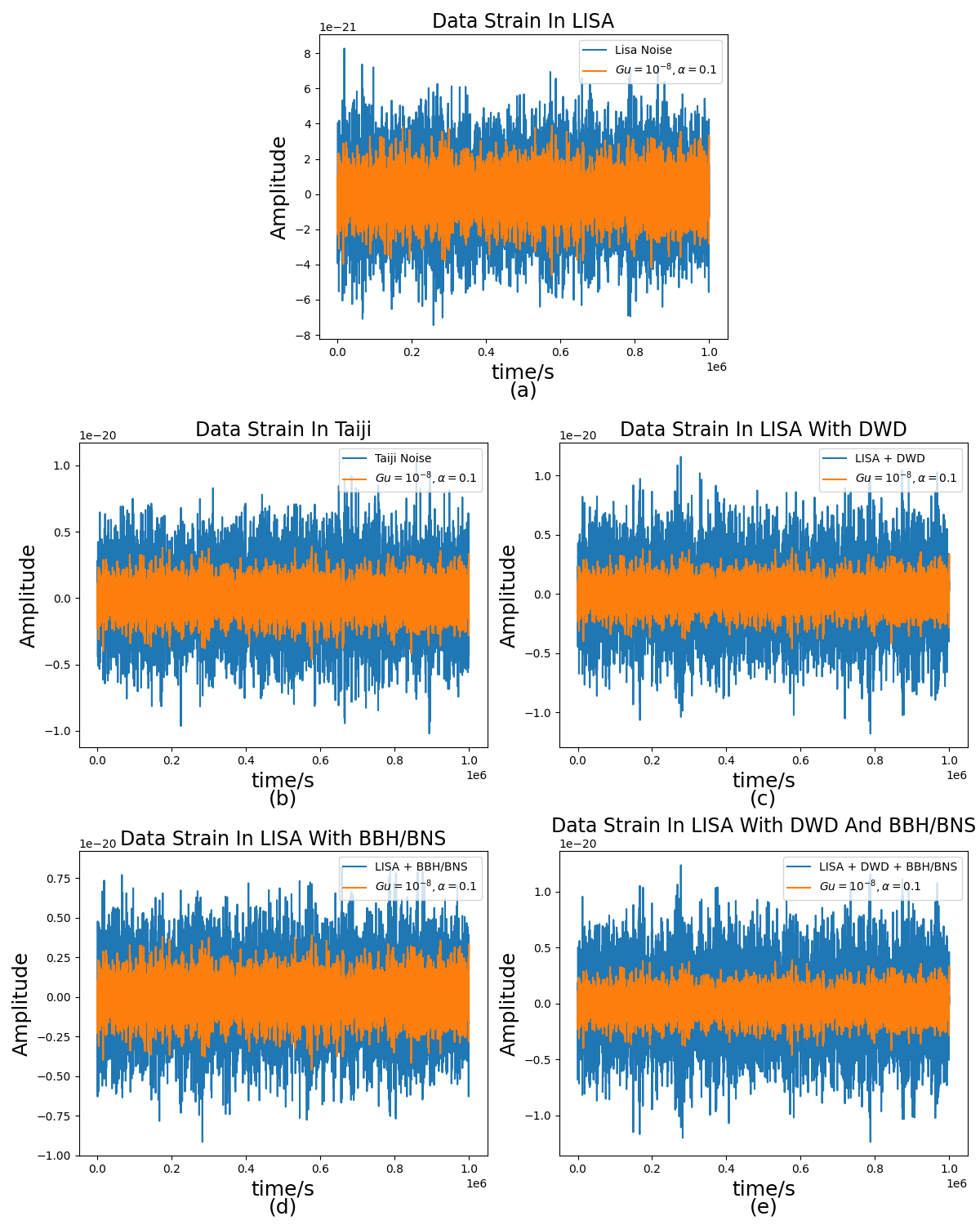} 
    \caption{\footnotesize Some data samples. (a) shows the instrumental noise of LISA and the SGWB in the time-domain data, while (b) illustrates the instrumental noise of Taiji. (c) shows the data with double white dwarf (DWD) noise, (d) shows the data with inspiral BBH/BNS based on the observations of LIGO and Virgo, and (e) shows the data with both types of foreground noise.}
    \label{}
\end{figure}
\begin{figure*}[ht]
    \centering
    \includegraphics[scale=0.2]{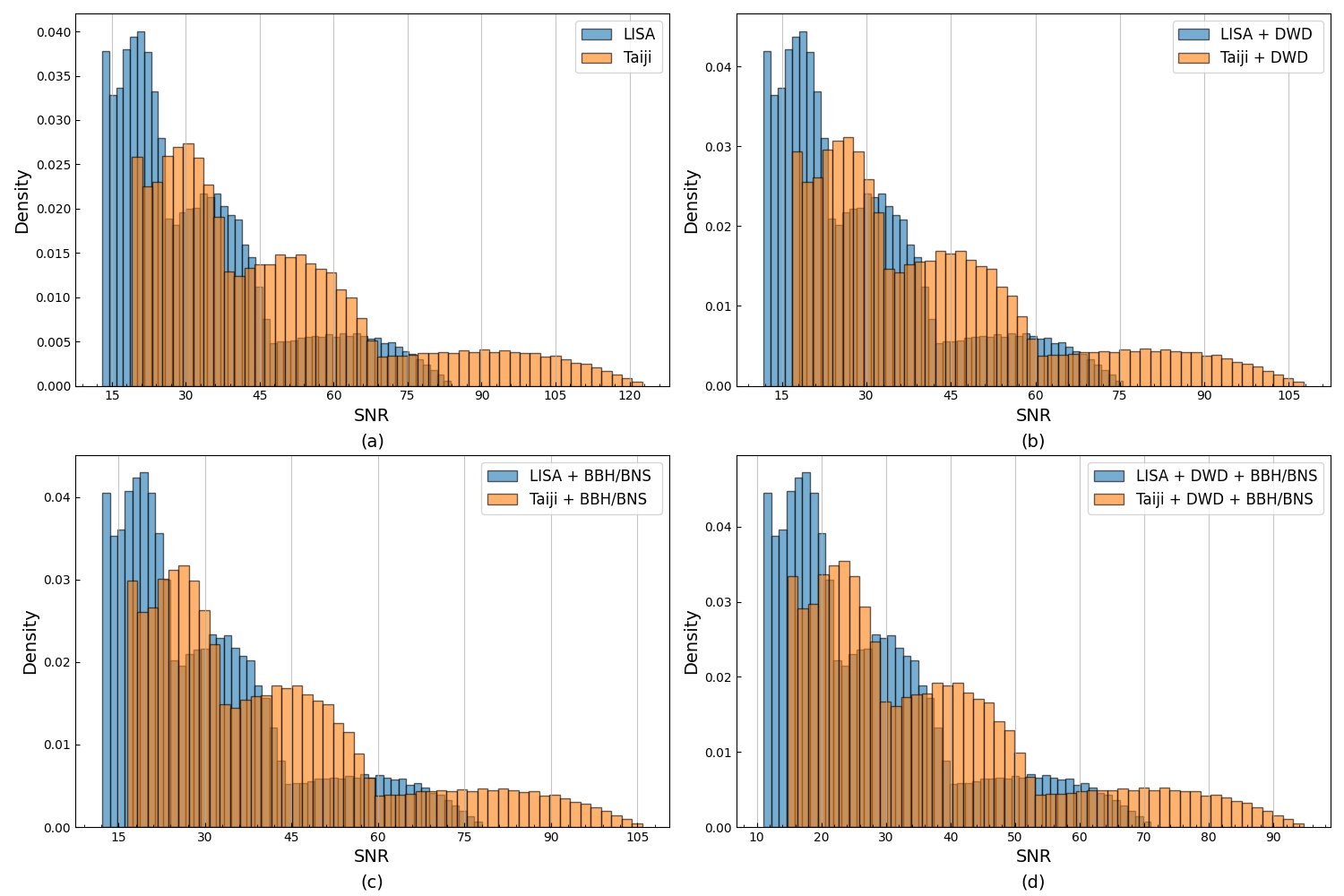} 
    \caption{\footnotesize The SNR distributions of our samples. The blue column represents the SNR for the LISA detector, while the orange column represents the SNR for the Taiji detector. (a) shows the SNR with only the detector's intrinsic noise; (b) shows the data with DWD foreground noise on top of (a); (c) shows the data with inspiral BBH/BNS based on the observations of LIGO and Virgo foreground noise on top of (a); and (d) shows the data with both DWD and inspiral BBH/BNS based on the observations of LIGO and Virgo foreground noises.}
    \label{}
\end{figure*}

In Figure 2, we present example samples of the generated data. As shown, the SGWB behaves like noise in the time domain, which complicates the process of signal identification. In Figure 3, we illustrate their SNR distributions. For the same data, the minimum SNR of LISA is 13, while Taiji achieves a minimum SNR of 19. Similarly, the maximum SNR for Taiji is 122.2, whereas LISA only reaches a maximum SNR of 83. This difference arises because Taiji, with its longer arm lengths compared to LISA, has lower instrumental noise, resulting in a higher SNR. Additionally, the SNR distribution for LISA is more concentrated. However, the presence of foreground noise adds complexity and chaos to the data, further increasing the difficulty of signal identification. The SNR is calculated using the formula provided in \cite{haris2018identifying}.
\begin{align}
\rho  = 2{\left( {\int_{{f_{\min }}}^{{f_{\max }}} {\frac{{{{\left| {h\left( f \right)} \right|}^2}}}{{S\left( f \right)}}df} } \right)^{\frac{1}{2}}},
\end{align}
where $S\left( f \right)$ is the PSD of the noise and $h\left( f \right)$ is the signal.

\section{Deep Learning}

\begin{figure*}[ht]
    \centering
    \includegraphics[scale=0.4]{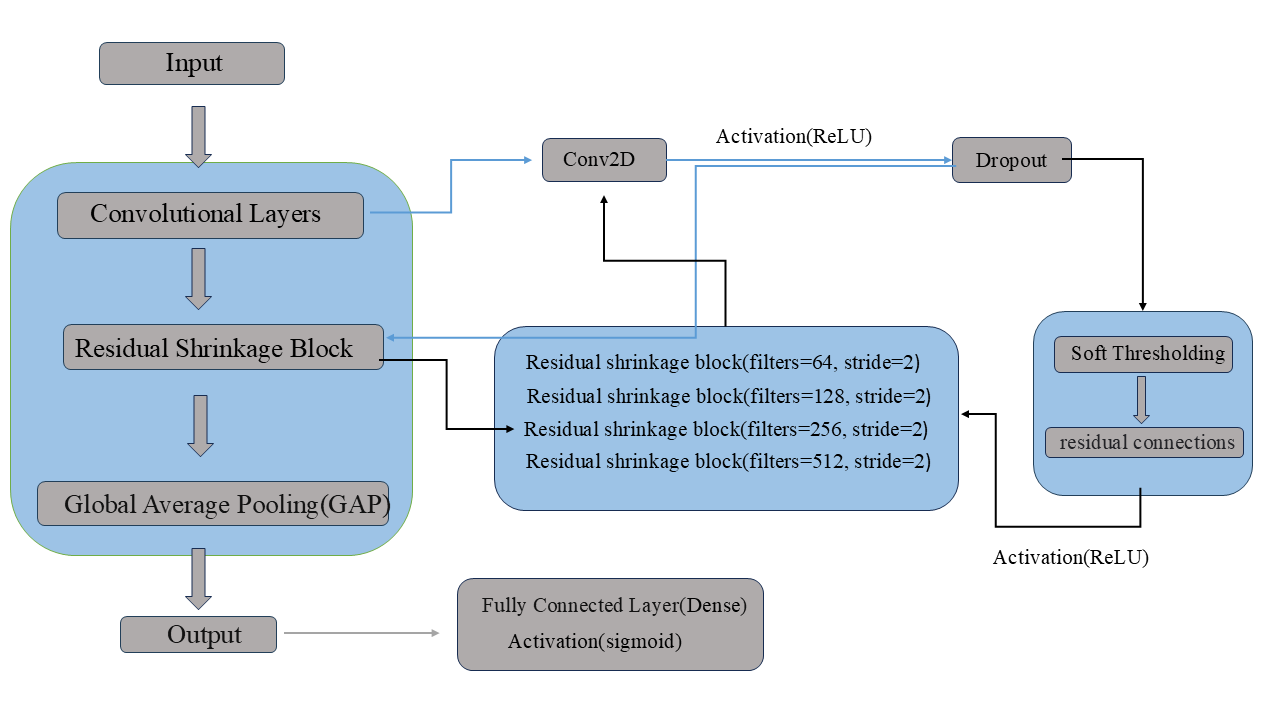} 
    \caption{\footnotesize The model architecture Schematic representation of our machine learning model.We employed a ResNet architecture and incorporated Dropout and regularization at appropriate stages. An initial convolutional layer is added before the residual blocks to activate the input data and extract preliminary features. The residual block consists of four stacked Residual Shrinkage Blocks (RSBs), each containing two convolutional layers. Dropout is applied between the two convolutional layers, followed by soft-thresholding to suppress noise and highlight important features. A global average pooling layer is then used to compress the high-dimensional features into low-dimensional vectors, facilitating subsequent classification. Finally, a fully connected Dense layer with a sigmoid activation function outputs the binary classification probabilities.}
    \label{}
\end{figure*}

We employed a Residual Shrinkage Network (ResNet) for this learning task, incorporating Dropout and regularization to mitigate overfitting \cite{zhao2019deep, li2021image, liu2024improved}. The ResNet is a deep learning model that combines residual networks with soft thresholding, aiming to enhance robustness against high-noise data. It is particularly well-suited for processing time-domain signal data. Figure 4 illustrates the streamlined workflow of the network we utilized. The architecture of our network is characterized by the following key features: residual connections \cite{szegedy2017inception, he2016deep, yu2017dilated}, soft thresholding \cite{donoho1995noising, joy2013denoising, bredies2008linear}, regularization \cite{girosi1995regularization, tian2022comprehensive}, Dropout \cite{srivastava2014dropout, tinto1975dropout}, and multi-scale feature extraction \cite{pauly2003multi}.

Each residual block includes a skip connection \cite{orhan2017skip, wang2022uctransnet}, which directly adds the input to the output. Skip connections help alleviate the vanishing gradient problem and enable the network to more effectively learn identity mappings. Soft thresholding is integrated into each residual block, where an adaptive threshold is computed based on the global average pooling of the mean absolute value of the input features. This operation effectively filters out noise, significantly enhancing signal recognition performance, particularly in low signal-to-noise ratio (SNR) scenarios.

Our loss function is Focal Loss \cite{mukhoti2020calibrating}\cite{li2020generalized}, a weighted cross-entropy loss function designed to enhance the model's emphasis on samples that pose greater classification challenges, which is 
\begin{align}
{L_{Focal}} =  - \frac{1}{N}\sum\limits_{i = 1}^N {{\alpha ^{\left( i \right)}}}  \cdot {\left( {1 - p_t^{\left( i \right)}} \right)^\gamma } \cdot \log \left( {p_t^{\left( i \right)}} \right)\,
\end{align}
where, $\alpha^{(i)}$ represents the weighting coefficient for each sample, and $p_t^{(i)}$ denotes the predicted probability for the sample, expressed as follows:
\begin{align}
{\alpha ^{\left( i \right)}} = {y_i}{\alpha ^*} + \left( {1 - {y_i}} \right)\left( {1 - {\alpha ^*}} \right),
\end{align}
and
\begin{align}
p_t^{\left( i \right)} = {y_i}{\hat y_i} + \left( {1 - {y_i}} \right)\left( {1 - {{\hat y}_i}} \right),
\end{align}
where, $y_i$ represents the ground truth label, $\hat{y}_i$ denotes the predicted probability for the positive class, ${\alpha ^*}$ and $\gamma$ are hyperparameters, and $N$ indicates the total number of samples.

To mitigate the complexity of training and enhance overall efficiency, we employed a curriculum-based learning approach \cite{lisicki2020evaluating, hacohen2019power, bengio2009curriculum}, which emphasizes a learning sequence from simple to complex. Specifically, this method involves adjusting the order of training samples based on their difficulty, starting with simpler or more easily understood samples and gradually introducing more complex ones. This approach helps the model better learn intricate patterns, thereby improving both training efficiency and performance.

In our study, the dataset consists of 200,000 samples, with an equal number of positive and negative samples. The positive samples are sorted by SNR in descending order. Starting from the highest SNR, every 20,000 positive samples, along with an equal number of negative samples, are selected together to form a subset, resulting in five subsets. Each subset thus contains 20,000 positive samples and 20,000 negative samples. For each subset, 20\% of the samples are randomly selected as the validation set, while the remaining 80\% are used for training.

Correspondingly, the training process is divided into five stages. In the first stage, only the subset with the highest SNR is used for training. In the second stage, the subset with a slightly lower SNR is added to the training set. This progression continues through the third and fourth stages. Finally, in the fifth stage, the entire training dataset is used for training \cite{hacohen2019power, bengio2009curriculum}. This progressive training approach allows the model to initially learn fundamental features from high-SNR samples, gradually adapting to the complexities introduced by low-SNR samples, thereby enhancing overall learning performance. Table 1 presents some of the hyperparameters used in the network.

\begin{table}[ht]
\caption{The hyperparameters of the network.}
\centering
\begin{tabular}{|c|c|c|}
\hline
\label{tab:simple_table}

optimizer       &  Adam                                             \\ \hline
learning rate   &  0.0003/0.0003/0.0003/0.0001/0.0001               \\ \hline
loss            &  Focal Loss(${\alpha ^*}$=0.25/0.25/0.2/0.2/0.2)  \\ \hline
loss            &  Focal Loss($\gamma $=1/2/2/2/2)                  \\ \hline
epochs          &  10/20/20/30/30                                   \\ \hline
batch size      &  64/64/64/32/32                                   \\ \hline
\end{tabular}

\end{table}

\section{Results and Analysis}
An early stopping strategy was employed during the training process, which involved terminating the current phase of training and moving to the next phase when the loss showed minimal change over five consecutive epochs. In the first phase, the loss for all three models decreased rapidly to a low value. As each new phase introduced additional samples with lower SNR for training, a noticeable increase in loss occurred at the beginning of each phase. The curriculum learning strategy, which leverages high-SNR samples to establish a robust initial framework, ensures that the loss variations in subsequent phases remain moderate.

We analyzed the performance of individual detectors and joint detection in identifying the stochastic gravitational wave background separately. The results show that joint detection outperforms individual detectors, with Taiji achieving better results than LISA. When foreground noise is included, the detection accuracy decreases for both individual detectors and joint detection. However, joint detection still demonstrates superior performance compared to individual detectors. Additionally, in the cases with foreground noise, we analyzed how detection accuracy varies with SNR.

\subsection{Machine learning performance with  instrumental noise}
The ROC curve and confusion matrix are commonly used to evaluate the performance of machine learning models. In this section, we focus solely on the data containing the SGWB and instrumental noise. As shown in Figure 5, the ROC curve for the LISA detector has the smallest area, with an AUC value of 0.9453. In contrast, machine learning with the Taiji detector and the joint detection of both detectors achieved AUC values exceeding 0.98, demonstrating excellent detection performance. 

\begin{figure}[h!]
    \centering
    \includegraphics[scale=0.28]{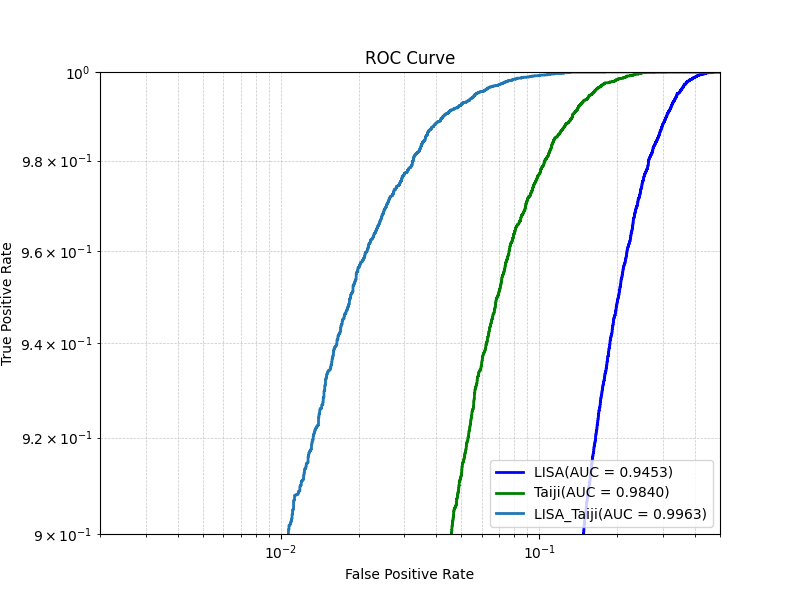} 
    \caption{\footnotesize The ROC curves for LISA, Taiji, and the joint detection of both detectors under instrument noise.}
    \label{}
\end{figure}

As shown in Figure 6, the majority of misclassifications involve incorrectly identifying signal-containing samples as noise-only samples, with this type of error occurring more than twice as frequently as the other type of misclassification. This is primarily due to the fact that the physical behavior of the SGWB closely resembles that of noise. The detection accuracies for LISA, Taiji, and joint detection are 87.60\%, 94.11\%, and 97.31\%, respectively. It is clear that joint detection outperforms the individual detectors in SGWB identification.

\begin{figure*}[ht]
    \centering
    \includegraphics[scale=0.33]{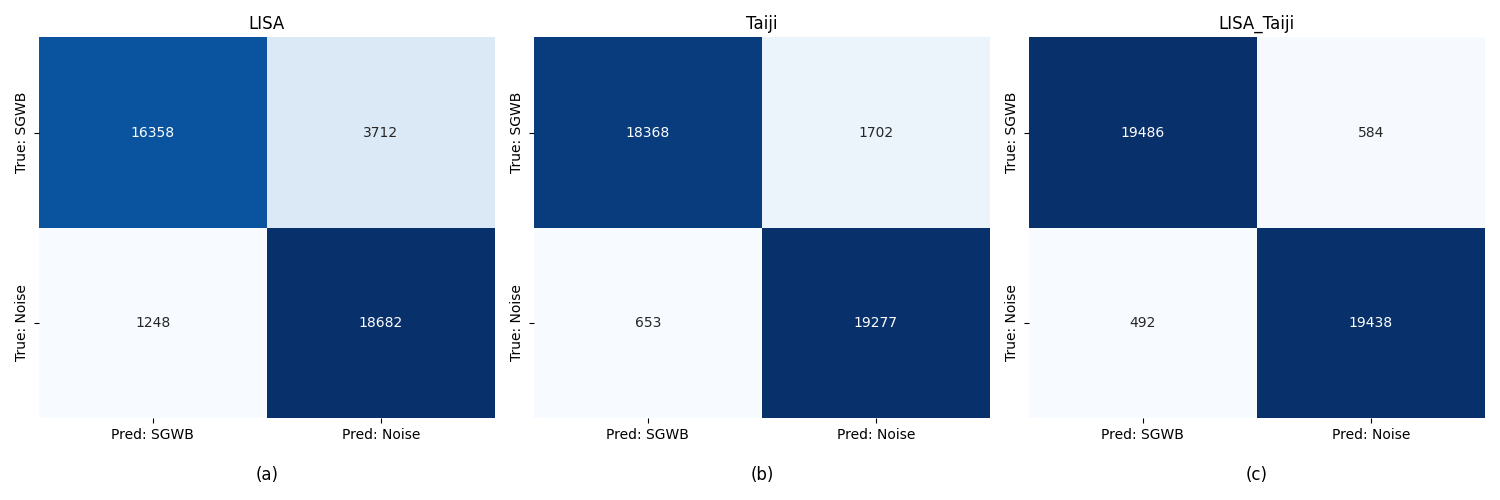} 
    \caption{\footnotesize The confusion matrices to the results of LISA, Taiji, and joint detection, with detection accuracies of 87.60\%, 94.11\%, and 97.31\%, respectively. }
    \label{}
\end{figure*}

As shown in Figure 7, for LISA, a significant increase in loss is observed in the third phase; for Taiji, this increase becomes evident in the fourth phase; and for joint detection, the increase is only noticeable in the fifth phase. Furthermore, the final loss for joint detection is considerably lower than that for the individual detectors.

\begin{figure}[h!]
    \centering
    \includegraphics[scale=0.26]{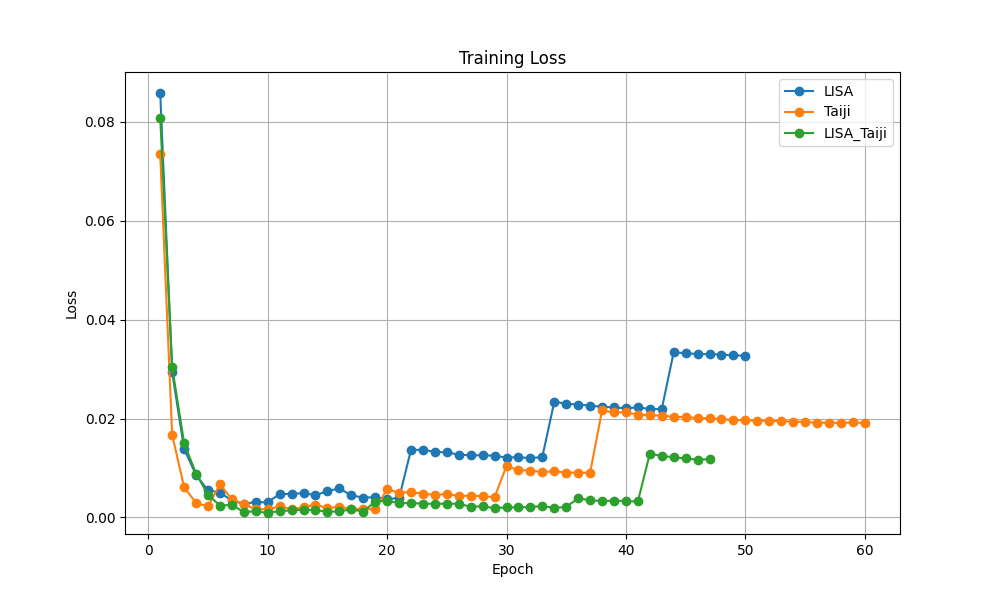} 
    \caption{\footnotesize The loss variations with epochs for the LISA, Taiji, and joint detection models. At lower SNRs, the loss for joint detection is consistently smaller than that of the individual detectors. The significant increase in loss at the beginning of a new phase occurs due to adding the lower SNR samples}
    \label{}
\end{figure}

By analyzing the ROC curves, confusion matrices, and loss reduction plots, we conclude that joint detection outperforms individual detectors in identifying and detecting SGWB signals. Consequently, joint detection is likely to be a more effective and feasible approach for SGWB signal detection.

\subsection{Machine learning performance with foreground noise}
In our previous analysis, we concluded that joint detection provides higher accuracy compared to using a single detector. Therefore, the analysis in this section is based on joint detection. We further investigated the impact of SNR ranges on machine learning performance, considering the DWD and inspiraling BBH/BNS foreground noise based on the observations of LIGO and Virgo. Specifically, we use several datasets with different SNR ranges: 200,000 samples (SNR range: 11 to 71), 180,000 samples (SNR range: 13.5 to 71), 160,000 samples (SNR range: 15.8 to 71), 140,000 samples (SNR range: 18 to 71), and 120,000 samples (SNR range: 20 to 71). Since the SNR for the same positive sample can be correspondingly obtained in both the Taiji and LISA detectors, we refer to the SNR values under the LISA detector for joint detection throughout this analysis, in order to avoid any confusion.

\begin{figure}[h!]
    \centering
    \includegraphics[scale=0.28]{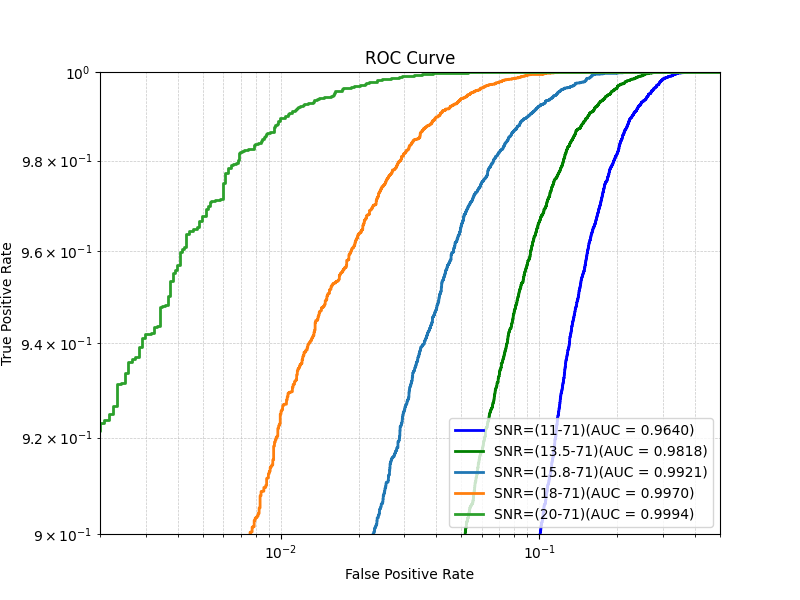} 
    \caption{\footnotesize ROC curve for different SNR.}
    \label{}
\end{figure}

\begin{figure*}[ht]
    \centering
    \includegraphics[scale=0.3]{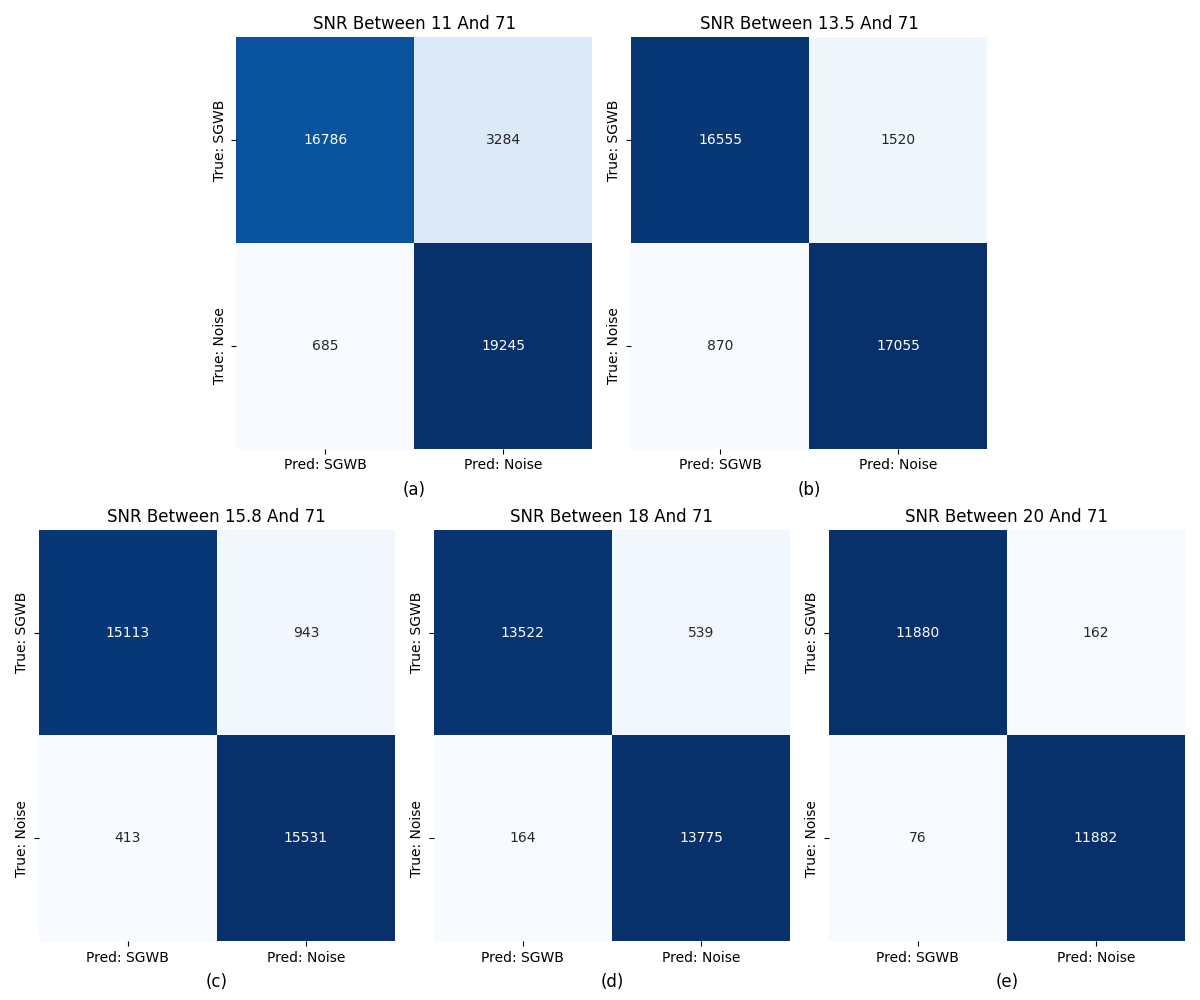} 
    \caption{\footnotesize Confusion matrix for different SNR. (a) shows a SNR between 11 and 71 with 90.07\% accuracy. (b) shows a SNR between 13.5 and 71 with 93.36\% accuracy. (c) shows a SNR between 15.8 and 71 with 95.76\% accuracy. (d) shows a SNR between 18 and 71 with 97.45\% accuracy. (e) shows a SNR between 20 and 71 with 97.45\% accuracy.}
    \label{}
\end{figure*}

As shown in Figures 8 and 9, when the recognition accuracy exceeds 95\%, the corresponding AUC of the ROC curve consistently remains above 0.99. Figure 8 underscores the importance of selecting appropriate SNR ranges for evaluating model performance, and shows how AUC values vary across different SNR intervals. Notably, within the SNR range of 15.8 to 71, the AUC values consistently exceed 0.99, indicating superior model performance in this range. Figure 9 further illustrates the recognition accuracy across different SNR ranges. Starting from the lowest SNR range, the recognition accuracies are 90.07\%, 93.36\%, 95.76\%, 97.45\%, and 99.00\%, respectively. These results demonstrate that an SNR range of 15.8 to 71 provides a solid foundation for feature learning, thereby enhancing both recognition performance and generalization capability.

However, this does not imply that all samples with SNRs greater than 15.8 can be classified correctly. While high-SNR samples provide a solid foundation for feature learning and classification, they may still be misclassified due to other factors, such as noise complexity or subtle variations in signal characteristics. Furthermore, samples with SNRs below 15.8 pose a greater challenge to the model, as their signal features become more indistinguishable from noise, resulting in a higher misclassification rate.

\begin{figure}[h!]
    \centering
    \includegraphics[scale=0.4]{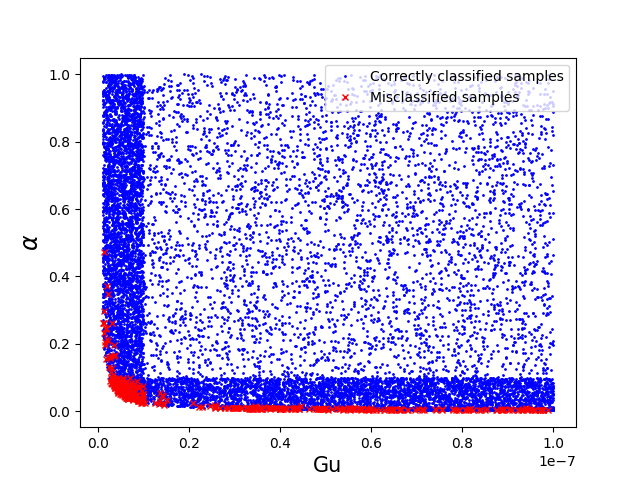} 
    \caption{\footnotesize The classification in the parameter space. Blue dots represent the samples correctly classified by the model, while red crosses represent the misclassified samples, which corresponds to Figure 9(c).}
    \label{}
\end{figure}

These results indicate that within the SNR range of 15.8–71, even with the presence of foreground noise, machine learning with joint detection can still effectively identify SGWB signals in the data. As shown in Figure 10, after adding the two types of foreground noise, only a small fraction of the samples remain unrecognized. For parameter ranges where $G{\rm{u}} \ge 2 \times {10^8}$ and $\alpha  \ge 0.1$, the samples can be correctly classified by our machine learning model. This suggests that SGWB signals from cosmic strings with these parameter values may be detectable using machine learning in future space-based gravitational wave joint detection efforts.

With this SNR range (15.8–71), we evaluated the performance of machine learning with three types of foreground noise: only DWD noise, only inspiral BBH/BNS noise based on LIGO and Virgo observations, and a mixture of both noise types. The total number of samples was reduced from 200,000 to 160,000, with positive and negative samples equally distributed. Figures 11 and 12 show that the AUC of the ROC curves for all three cases exceeds 0.99, and the accuracy remains above 95\%. The best results are from inspiraling BBH/BNS noise based on LIGO and Virgo observations with 99.07\% accuracy. For the case with only DWD foreground noise, the accuracy reaches 97.56\%. Even with the simultaneous addition of both types of foreground noise, the model still achieves an accuracy of 95.76\%.

\begin{figure}[h!]
    \centering
    \includegraphics[scale=0.28]{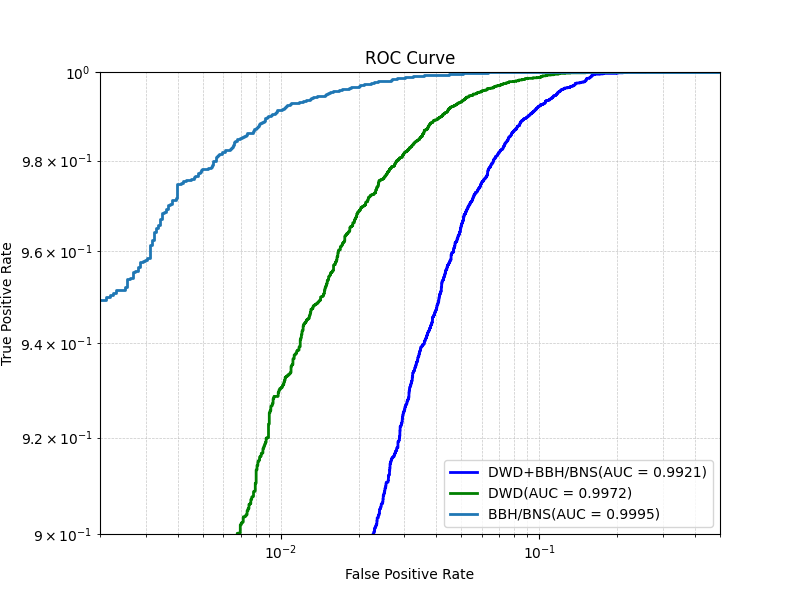} 
    \caption{\footnotesize ROC curve with different foreground noises in the joint detection. The analysis includes the simultaneous injection of the aforementioned two types of foreground noise, using samples with an SNR range of 15.8 to 71.}
    \label{}
\end{figure}

\begin{figure*}[ht]
    \centering
    \includegraphics[scale=0.3]{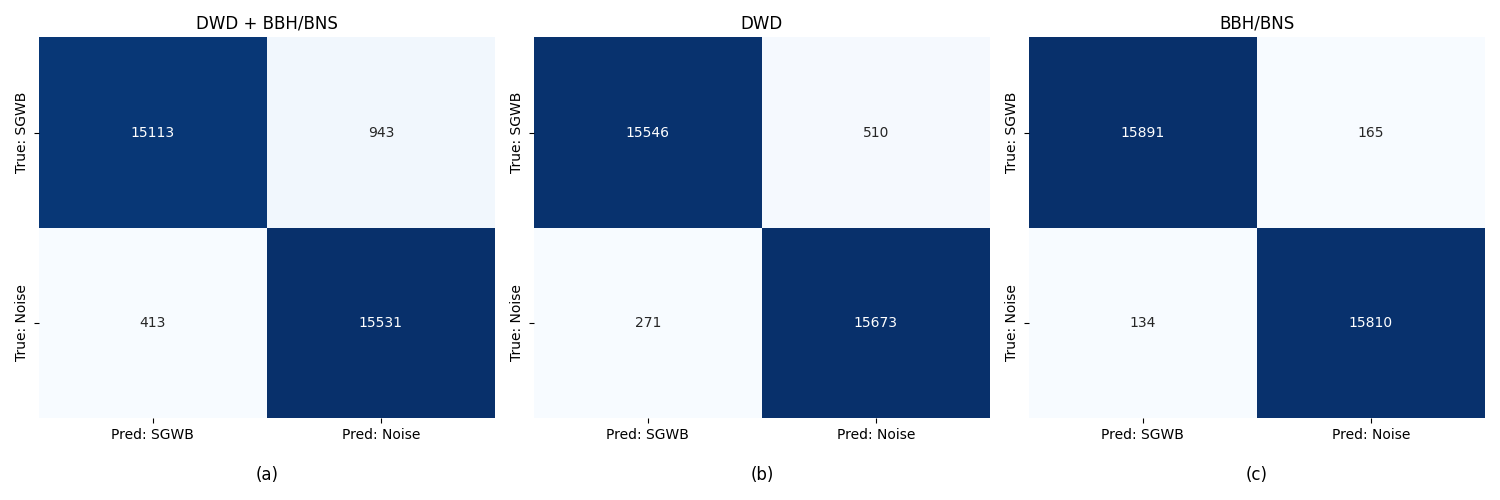} 
    \caption{\footnotesize Confusion matrix for different foreground noise. The sample selection is consistent with Figure 11. (a) shows data with DWD foreground noise and inspiral BBH/BNS noise based on LIGO and Virgo observations. (b) shows data with DWD foreground noise. (c) shows data with inspiral BBH/BNS noise based on LIGO and Virgo observations.}
    \label{}
\end{figure*}

In summary, the results demonstrate that restricting the SNR range (e.g., selecting samples with SNRs between 15.8 and 71) significantly enhances the model's overall performance. This underscores the critical role of high-SNR samples in improving model accuracy, while highlighting the necessity of future efforts to refine algorithms and models to better handle low-SNR samples in complex signal-noise environments.

\section{Summary and Discussion}

In our research, the computational configuration consists of an NVIDIA RTX 3090 GPU and an Intel(R) Xeon(R) Silver 4214R CPU (2.40 GHz, 48-thread processor). We employed full analytical approximations to simulate the SGWB generated by cosmic strings and modeled the instrumental noise of both the LISA and Taiji detectors based on their sensitivity curves. By leveraging residual shrinkage networks and curriculum learning strategies, we developed a robust model and demonstrated that multi-detector joint observations significantly enhance the detection of SGWB signals.

In traditional machine learning and deep learning training, training samples are typically shuffled, and the model encounters samples of varying difficulty throughout the process, with these difficulties being uniformly distributed. In contrast, curriculum learning mimics human learning by gradually guiding the model through the learning process according to the difficulty of the samples, enabling it to acquire knowledge more efficiently.

These results suggest that machine learning holds great promise for the effective detection of SGWB signals by space-based gravitational wave detectors. Furthermore, combining multiple detectors could further improve the performance of machine learning models, enabling more efficient and accurate identification of SGWB signals.

Moreover, foreground noise presents a significant challenge for space-based detectors, as many of the received data streams are likely to be substantially affected by such noise. To address this, we incorporated foreground noise into our study, including contributions from DWD and inspiraling BBH/BNS, based on the observations of LIGO and Virgo. By analyzing the model's performance across different SNR ranges, we found that our machine learning model can effectively handle the challenges posed by foreground noise, providing a solid foundation for future SGWB signal detection efforts.

The discussions presented here offer a meaningful attempt and valuable reference for future space-based gravitational wave data analysis powered by machine learning. Certainly, the observational data acquired by future gravitational wave detectors will exhibit substantially greater complexity than the simulated data analyzed in this study. The refined cosmic string models will impose more stringent constraints on gravitational wave production\cite{hindmarsh2023multi,kume2024revised}, thereby presenting significant challenges for the application of machine learning techniques in both data analysis and theoretical model interpretation. We plan to research it in the future work.

\section*{Acknowledgements}
This work was supported by the National Key Research and Development Program of China (Grant No. 2021YFC2203004), the National Natural Science Foundation of China (Grant No. 12347101), the Natural Science Foundation of Chongqing (Grant No. CSTB2023NSCQ-MSX0103).

\nocite{*}
\bibliography{bibliography}

\end{document}